\documentclass[aps,showpacs,prd,tightenlines,twocolumn,10pt,nofootinbib]{revtex4}

\usepackage[latin1]{inputenc}
\usepackage{bbm}
\usepackage{graphicx}
\usepackage{epsfig}
\usepackage{subfigure}

\usepackage{color}

\newcommand{\be}{\begin{eqnarray}} % only untightened
\newcommand{\ee}{\end{eqnarray}}

\newcommand{\bmp}{\noindent\begin{minipage}{16cm}}
\newcommand{\emp}{\end{minipage}\vskip 7mm} % 7mm untightened
\def\lsim{\mathrel{\raise.3ex\hbox{$<$\kern-.75em\lower1ex\hbox{$\sim$}}}}
\def\gsim{\mathrel{\raise.3ex\hbox{$>$\kern-.75em\lower1ex\hbox{$\sim$}}}}

\newcommand{\intron}[1]{}%{#1}

\begin{document}

%%%%%%%%%%%%%%%%%%%%%%%%%%%%%%%%%%%%%%%%%%%%%%%%%%%%%%%%%%%%%

\title{The WIMP of a Minimal Technicolor Theory}

\author{Kimmo {\sc Kainulainen}$^1$, Kimmo {\sc Tuominen}$^{1,2}$ and Jussi 
{\sc Virkaj\"arvi}$^{1,2}$}
\affiliation{$^1$Department of Physics, University of Jyv\"askyl\"a, Finland \\
$^2$Helsinki Institute of Physics, University of Helsinki, Finland}
%\date{\today, \currenttime}

%%%%%%%%%%%%%%%%%%%%%%%%%%%%%%%%%%%%%%%%%%%%%%%%%%%%%%%%%%%%%

\begin{abstract}
We consider the possibility that a massive fourth family neutrino, predicted by a recently proposed minimal technicolor theory, could be the source of the dark matter in the universe. The model has two techniflavors in the adjoint representation of an SU(2) techicolor gauge group and its consistency requires the existence of a fourth family of leptons. By a suitable hypercharge assignement the techniquarks together with the new leptons look like a conventional fourth standard model family. We show that the new (Majorana) neutrino $N$ can be the dark matter particle if $m_N \sim 100-500$ GeV and the expansion rate of the Universe at early times is dominated by an energy component scaling as $\rho_\phi \sim a^{-6}$ (kination), with $\rho_\phi/\rho_{\rm rad} \sim 10^{-6}$ during the nucleosynthesis era.


\end{abstract}

%%%%%%%%%%%%%%%%%%%%%%%%%%%%%%%%%%%%%%%%%%%%%%%%%%%%%%%%%%%%%%%%%%%%%%%%%%%%%

\maketitle

%%%%%%%%%%%%%%%%%%%%%%%%%%%%%%%%%%%%%%%%%%%%%%%%%%%%%%%%%%%%%%%%%%%%%%%%%%%%%

\section{Introduction}

According to the presently favored cosmological model the universe is to a high precision flat with baryonic, dark matter and dark energy densities of  $\Omega_{\rm b,0} \simeq 0.04$, $\Omega_{\rm m,0} \simeq 0.20$ and $\Omega_{\Lambda} \simeq 0.76$ respectively~\cite{Spergel:2006hy}. The true nature of neither the dark matter (DM), nor the dark energy (DE) is known at present, although many candidates for the particle DM have been proposed, in particular in the context of supersymmetry~\cite{Lahanas:2006mr}. Dark energy is even more enigmatic, as even the theoretical framework to explain its origin remains speculative. Barring the simplest - and the least attractive - possibility that the $\Omega_{\Lambda}$ is a constant, the time evolution of the dark energy is of potential interest for cosmology~\cite{peeblesratra,SNIa}.

Dark energy may also affect the very early cosmology. Indeed, many models for a dynamical dark energy, including quintessence~\cite{quintessence}and scalar tensor gravity theories~\cite{STGM}, can exhibit an early phase where the energy of the universe is dominated by the kinetic energy of a rolling scalar field. During this {\em kination} phase~\cite{Joyce:1996cp} the energy-density of the universe scales as $\rho \approx \rho_\phi \sim a^{-6}$.  To ensure the success of the big bang nucleosynthesis the universe must be radiation dominated and the ratio $(\rho_\phi/\rho_{\rm rad})_* \equiv r$ must be a small number during the BBN era at $T \simeq 1$ MeV.  However, as the kinetic energy scales more strongly than does the radiation $\rho_{\rm rad} \sim a^{-4}$, it can dominate the expansion at times earlier than BBN. Other physical phenomena that occur before BBN may thus be strongly affected by the dark energy induced kination. This includes in particular the decoupling of weakly interacting dark matter particles (WIMPs). In a normal expansion history WIMPs with weak interaction strenghs decouple at $T_{\rm dec} \sim m_N/30-m_N/20$, while their number density is scaling as $\sim e^{m_N/T}$.  Increasing expansion rate due to kination leads to an earlier decoupling and hence to a potentially much larger final DM-density than in the standard cosmology~\cite{Michael}. The effect has been studied for generic WIMPs in the context of a quintessence cosmology in ref.~\cite{Salati:2002md} and for a scalar tensor gravity in ref.~\cite{Catena:2004kw}. 
In this context it is interesting to observe that a recently discovered technicolor model~\cite{Sannino:2003xe, Sannino:2004qp}, originally introduced as an alternative theory for the electroweak symmetry breaking, predicts the existence  a heavy neutrino with normal weak interactions and with a mass in the range $m_N \sim 100-500$ GeV. Under standard expansion history the mass density of such a neutrino would be too small to be of interest~\cite{Enqvist:1988we}, but as we will show here, in a generic dark-energy cosmology with an early kination phase $N$ becomes a viable DM-candidate. 

Our cosmological model contains three essential parameters: $\Omega_N \equiv \Omega_{\rm m,0}(m_N,m_H,r)$, where $m_N$ and $m_H$ are the masses of the neutrino and of the effective Higgs excitation and $r$ is the ratio of the dark energy density to radiation energy density during nucleosynthesis. Neutrino mass $m_N$ and $r$ set the overall scale of the interaction and the expansion rates and the Higgs boson mass also plays a role, because the s-channel $N\bar N$ annihilation through a Higgs boson exchange to $W$-bosons displays a prominent peak in the cross section, reducing $\Omega_N$ strongly when $m_N \simeq m_H/2$. We find that $\Omega_N \approx 0.23$ can be arranged for a wide range of masses $m_N = 100-500$ GeV, given that the dark energy component at $T \sim 1$ MeV is of order $r \sim 10^{-4}$ (Dirac case) and $r \sim 10^{-6}$ (Majorana case) consistent with the nucleosynthesis. The constraints from direct searches~\cite{Akerib:2004fq} are rather strong for the Dirac case, and a Dirac-$N$ is ruled out as a dark matter unless the local WIMP density is roughly by an order of magnitude smaller than the conventional estimate.  Majorana neutrinos are much less constrained by the direct WIMP searches however, and turn out to be a good dark matter candidate in our model.

This paper is organized as follows: we will introduce the basic technicolor model and compute the precision electroweak data constraints for both the Dirac and the Majorana case in section II. We show that these constraints can be satisfied in both cases when the fourth doublet is such that the neutrino is the lighter of the doublet members, roughly with $m_L \gsim 2m_N$ and $m_N \gsim 100$ GeV, consistent with cosmology. In section III we compute $\Omega_N$ as a function of $m_N$ and $r$, and estimate the limits on masses set by the direct cryogenic searches, and we present our conclusions and outlook in section IV.

\section{Minimal walking technicolor and a fourth family of leptons}

Technicolor (TC) and the associated extended technicolor (ETC) theories were invented a long time ago to explain the mass patterns of the standard model gauge bosons and fundamental fermions~\cite{TC,Hill:2002ap}. Early TC models based on straightforward extrapolation from a QCD-like strongly interacting technicolor sector had several problems, including flavor changing neutral currents and unwanted additional light pseudo-Goldstone bosons. 
These problems were partly addressed in walking technicolor theories where the technicolor gauge coupling evolves slowly due to a near-conformal behavior. However, many fundamental techniflavors were needed to achieve this limit, and this led to the oblique $S$ parameter $S\approx {\cal O}(1)$ contrary to the observed value $S\approx 0$.

These conclusions only hold for theories with matter multiplets in the fundamental representation however, and can change when one uses higher fermion representations for techniquarks~\cite{Lane:1989ej,Corrigan:1979xf}. Along these lines, a minimal model of a walking techicolor was proposed~\cite{Sannino:2004qp}, which utilizes the recently discovered connections between supersymmetric gauge theories and non-supersymmetric gauge theories with matter in two index tensor representations~\cite{Armoni:2003gp, Sannino:2003xe}. It was shown that with just two techniflavors in the two-index symmetric representation of the SU$_{\rm{T}}$(2) gauge group the theory is already close to the conformal window. In \cite{Hong:2004td,Dietrich:2005jn,Dietrich:2005wk} it was shown that this model is compatible with precision measurements and that the composite Higgs boson can be light, with mass of the order of few hundred GeV \cite{Dietrich:2005wk}. A particularly interesting property for cosmology of the two color theory is that in addition to the techniquarks a new lepton doublet is required to exist. 

In the minimal model to be studied here, the electroweak symmetry breaking is driven by the gauge dynamics of two Dirac fermions in the two-index symmetric representation of the SU$_{\rm{T}}$(2) gauge theory. For SU(2) group the two index symmetric representation coincides with the adjoint representation, and as was shown in \cite{Sannino:2004qp}
this model is (quasi)conformal with {\it just} one doublet of techinfermions \cite{Sannino:2004qp}. To be specific, we introduce the techni-fermions as follows:
\be
  T_L^{\{C_1,C_2 \}} &=&
    \left(\begin{array}{l}
          U^{\{C_1,C_2 \}}
       \\ D^{\{C_1,C_2 \}}
          \end{array}
    \right)_L \ ,
\nonumber \\
  T_R^{\{C_1,C_2\}} &=&
    \left(U_R^{\{C_1,C_2\}},~ D_R^{\{C_1,C_2\}}
    \right) \ .
\nonumber
\ee
Here $C_i=1,2$ is the technicolor index and $T_{L(R)}$ denotes a doublet (singlet) with respect 
to the weak interactions. Since the two-index symmetric representation of SU$_{\rm{T}}$(2) is real, 
the global symmetry group ${\rm{SU}}_{\rm{L}}(2)\times {\rm{SU}}_{\rm{R}}(2)\times {\rm{U}}(1)$ is enhanced to ${\rm{SU}}(4)$, which breaks to ${\rm{SO}}(4)$ due to a formation of the fermion condensate $\langle\overline{U}U+\overline{D}D\rangle$. This pattern leads to the appearance of nine Goldstone bosons. Three of these will become the longitudinal components 
of the weak gauge bosons, and the low energy spectrum is expected to contain six quasi Goldstone bosons which receive mass through ETC interactions~\cite{Hill:2002ap,Lane:2002wv,Appelquist:2002me,Appelquist:2004ai}.

The weak interactions are affected by the global SU(2) Witten anomaly \cite{Witten:fp}, and hence the new weak doublets must be included in even numbers. In the model under consideration  the techniquarks are in the adjoint representation of SU(2), so that they correspond to three extra left doublets from the weak interactions point of view. As pointed out in \cite{Sannino:2004qp}, and implemented in detail in \cite{Dietrich:2005jn}, a simple way to cure this anomaly is to introduce at least one new lepton generation. While also other possibilities do exist \cite{Dietrich:2005jn}, we will here consider the hypercharge assignments of the techiquarks and new leptons such that these particles form just a fourth standard-model like family. Then the electric charges of the tecniquarks are as in the standard model: $Q(U)=+2/3$, $Q(D)=-1/3$, the charged lepton has charge -1 and the associated neutrino is neutral. Another anomaly free hypercharge assignment of this model under which a neutral technibaryonic dark matter candidate exists (but no leptonic ones) has been studied in \cite{Gudnason:2006ug, Gudnason:2006yj}.

\begin{figure*}[t]
\centering
  \subfigure{
  \includegraphics[width=7.6cm]{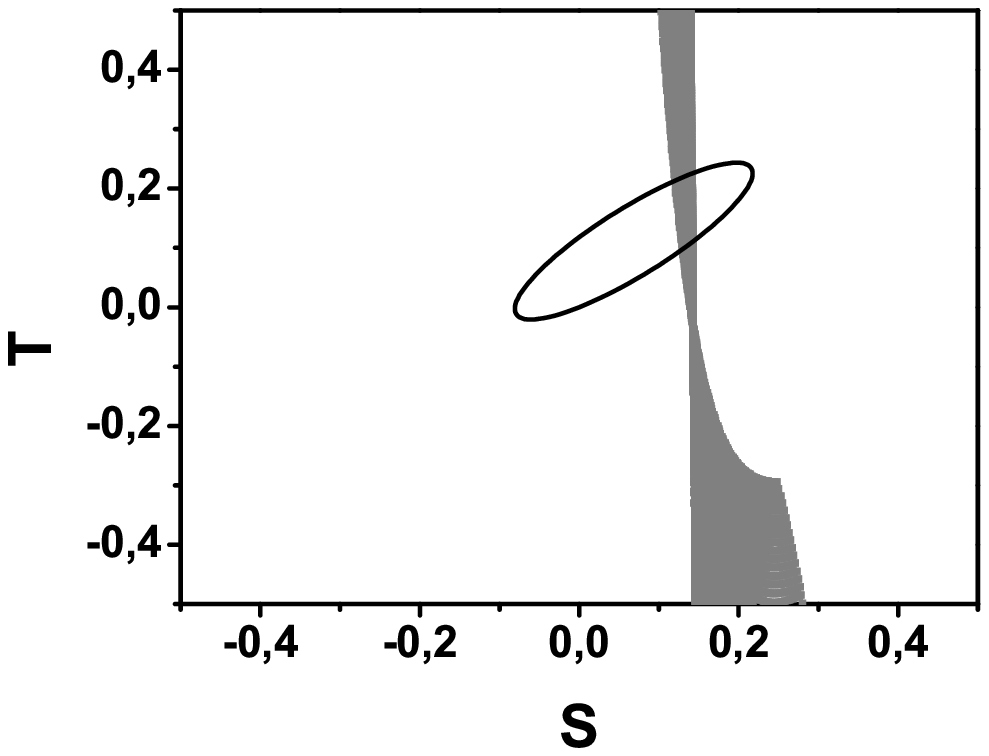}}
  \qquad
  \subfigure{\includegraphics[width=7.4cm]{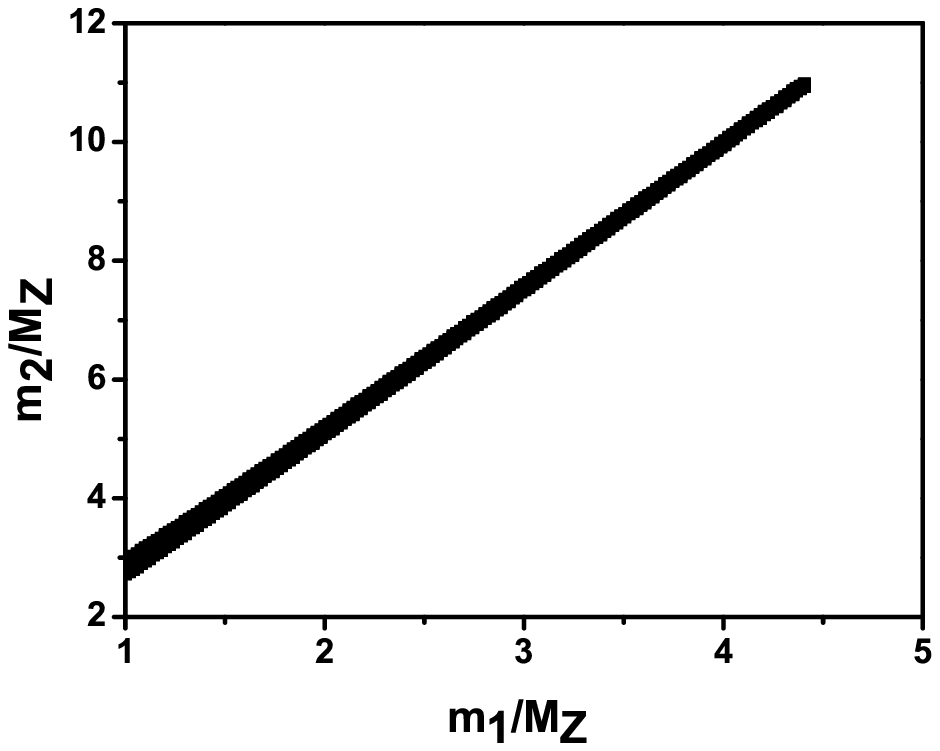}}
\vskip-0.4truecm
\caption{{\it Left Panel}: The shaded area corresponds to the accessible range for $S$ and $T$ with the masses of the new leptons ranging from $m_Z$ to $10 m_Z$. Only the 68\% confidence countour of the new global fit \cite{unknown:2005em} is shown. {\it Right Panel:} The shaded area shows the overlap region from the left panel as a function of the neutral (charged) lepton mass $m_1$ ($m_2$).}
\label{majorana_leptons}
\end{figure*}

In what follows we will denote the charged lepton by $\zeta$ and the associated neutrino by $N$.  Clearly the new lepton family must be sufficiently heavy and the model must not be at odds with the electroweak precision measurements. Just as in conventional standard model case, the charged lepton can only have a Dirac mass term while $N$ can have a Dirac, a Majorana or both mass terms. In the case of a pure Dirac-$N$ it has been shown~\cite{Dietrich:2005jn} that the electroweak precision data~\cite{Eidelman:2004wy,unknown:2005em} leads to constraints $M_N\geq M_Z$ and $M_\zeta\sim 2 M_N$\cite{Dietrich:2005wk}. 
Note that the results of this model are consistent with the more recent global fits \cite{unknown:2005em} within 68\% confidence limits over a wide range of masses of the fourth generation leptons. As discussed in \cite{Dietrich:2005jn} the possible additional Majorana mass term for the right handed neutrino 
ala Bertolini and Sirlin~\cite{Bertolini:1990ek} has a little impact on these results. 

The oblique corrections in the case of left handed neutrino with only a Majorana mass have not been discussed in previous work on this model, so we briefly present the results here. In this case the right handed neutrino is not needed at all, and the oblique corrections have been evaluated in \cite{Holdom:1996bn}. Including also the contribution from the techniquarks we have:
\begin{eqnarray}
  T &=& \frac{\pi}{s^2c^2M_Z^2}A(0) \\
  S &=& \frac{4\pi}{M_Z^2}[B(M_Z^2)-B(0)]+ \frac{1}{2\pi } \\
  U &=& \frac{4\pi}{M_Z^2}[A(M_Z^2)-A(0)],
\label{precision_majorana}
\end{eqnarray}
where $s=\sin\theta_W$, $c=\cos\theta_W$ and
\begin{eqnarray}
 A(q^2) &=&   
    2\Pi_{LL}^{N\zeta}(q^2)-\Pi_{LL}^{NN}(q^2)      
  -  \Pi_{LL}^{\zeta\zeta}(q^2)+\Pi_{LR}^{NN}(q^2) 
  \nonumber \\
 B(q^2) &=& 
     \Pi_{LL}^{NN}(q^2)-\Pi_{LL}^{\zeta\zeta}(q^2)
  - 2\Pi_{LR}^{\zeta\zeta}(q^2)-\Pi_ {LR}^{NN}(q^2).
\nonumber
\end{eqnarray}
Here $\Pi_{LL}^{ab}$ and $\Pi_{LR}^{ab}$ are the standard vacuum polarizations of left and right handed currents \cite{Peskin:1990pz}. The factor $1/(2\pi)$ in $S$ is the perturbative contribution of one generation of mass degenerate techniquarks. The non-perturbative contributions due to the near conformal behavior have been estimated to yield a reduction of the order of 20$\%$~\cite{Appelquist:1998xf}.  
Note that $T$ as defined above is proportional to the cutoff $\Lambda$ provided by the momentum dependence of the neutrino mass function. We take $\Lambda\approx 2 m_N$, see \cite{Holdom:1996bn}. The results in this case are summarized in Fig.~\ref{majorana_leptons}. From the left panel of the figure we see that also in this model there is a substantial overlap with the 68\% confidence contour of the precision data. This overlap region is depicted in terms of the masses of the charged and neutral new leptons in the right panel, from which we see that also in this case we have approximately $M_N\ge M_z$ together with $M_\zeta\sim 2 M_N$. 

Let us remark that the contribution to $T$-parameter can be negative in the Majorana case. This is shown in the left panel of Fig.\ref{majorana_leptons}, and it is to be contrasted with the Dirac case~\cite{Dietrich:2005wk} 
where $T$ is always positive.  However, the negative $T$ branch corresponds to the part of the parameter space where $N$ is heavier than $\zeta$, so that $N$ would not be a suitable dark matter candidate.  Negative contribution to $T$ might be of interest for future (ETC) model building however, because if the large value of the top quark is due to ETC interactions, these are likely to generate isospin violating effects around the TeV scale. Based on our results here, a natural framework to compensate for these seems to be to allow for some new Majorana particles. 

In this section we have established a technicolor model which naturally features a full fourth family of elementary fermions, which from the weak interaction viewpoint is just a replica of the conventional standard model generation. The leptons of the fourth generation, especially the neutrino, are very massive compared to the standard model ones and the masses are constrained by the electroweak precision data to lie above the direct observation limits. We have shown that consistency with the electroweak precision measurements allows the new neutrino to have either Dirac or Majorana mass, but constrains the relative mass splitting of the charged lepton and the associated neutrino qualitatively similarly in both cases. Assuming that the fourth generation neutrino does not mix with the standard model generations, a property which can be fully justified only if the more complete ultraviolet theory responsible for the masses and mixing of fundamental fermion fields is known, we now turn to the cosmological implications of such fourth generation neutrino.

\section{Relic Density}

We now compute the relic abundance $\Omega_N$ assuming that a quintessence-type dynamic dark energy component dominates the early stages of the expansion of the universe. The number density $n$ of heavy neutrinos can be computed from the usual Lee - Weinberg equation~\cite{Lee:1977ua}:
\begin{eqnarray}
\frac{\partial f(x)}{\partial x} 
        = \frac{\langle v\sigma\rangle 
         m_N^3 x^2}{H} (f^2(x)-f_{eq}^2(x)).  
\label{ecosmo1}
\end{eqnarray}
where we have introduced the scaled variables
\begin{eqnarray}
  f(x) \equiv \frac{n(x)}{s_E}, \quad {\rm and} \quad     
  x \equiv \frac{s_E^{1/3}}{m_N},                                                          
\end{eqnarray}
Here $s_E$ is the entropy density $s_E = ( \frac{2\pi^2}{45}g_{\ast s}(T))T^3$ at the temperature $T$ and $m_N$ is mass of the neutrino. Given the Hubble parameter $H$ and the average annihilation rate $\langle v\sigma \rangle$ the equation (\ref{ecosmo1}) can be solved numerically. For relatively light masses a direct integration is feasible, but for very heavy neutrinos we use an analytic approximation~\cite{Enqvist:1988dt}, which is accurate to about $5\%$. 

The early expansion rate of the universe in many quintessence-like or scalar tensor theories is dominated by the kinetic energy of the rolling scalar field. During this kination period the energy density in the scalar field scales like $\rho_\phi \sim a^{-6}$. This scaling behaviour obviously cannot be allowed during the BBN and so the $\rho_\phi$-component must be adjusted to be small during BBN. We take this into account by a simple parametrization which takes the Hubble parameter to the form
\begin{eqnarray}
  H = \bar H_0  
      \Big( \frac{x}{x_0}\Big)^2 \;
      \Big( h + r \, \big( \frac{x}{x_0}\big)^2 \Big)^{1/2}, 
\label{ecosmo2}
\end{eqnarray}
where $\bar H_0 = (8\pi\rho_{\rm rad,0}/3M_{\rm Pl}^2)^{1/2}$ is the Hubble expansion rate at the reference temperature $T=T_0 \equiv 1$ MeV in the standard cosmology, $r \equiv \rho_{\phi,0}/\rho_{\rm rad,0}$ is the ratio between dark energy and radiation energy densities at $T=T_0$ and $h \equiv (g_{*}(T)/g_*(T_0)) (g_{s*}(T)/g_{s*}(T_0))^{4/3}$, where $g_*(T)$ and $g_{*s}(T)$ denote the number of energy- and entropy degrees of freedom at the temperature $T$, respectively. 
For the averaged cross section we use the integral expression~\cite{Gondolo:1990dk}
\begin{eqnarray}
   \langle v \sigma \rangle &=& 
       \frac{1}{8m_N^{4}TK^{2}_2(\frac{m_N}{T})}
 \times\\ \nonumber
 && \phantom{han}\times
      \int_{4m_N^2}^{\infty}ds
                        \sqrt{s}(s-4m_N^2)K_1(\frac{\sqrt{s}}{T})
                        \sigma_{\rm tot}(s)
\label{ecosmo3}
\end{eqnarray}
\begin{figure*}[t]
\centering
  \subfigure{\includegraphics[width=6.47cm]{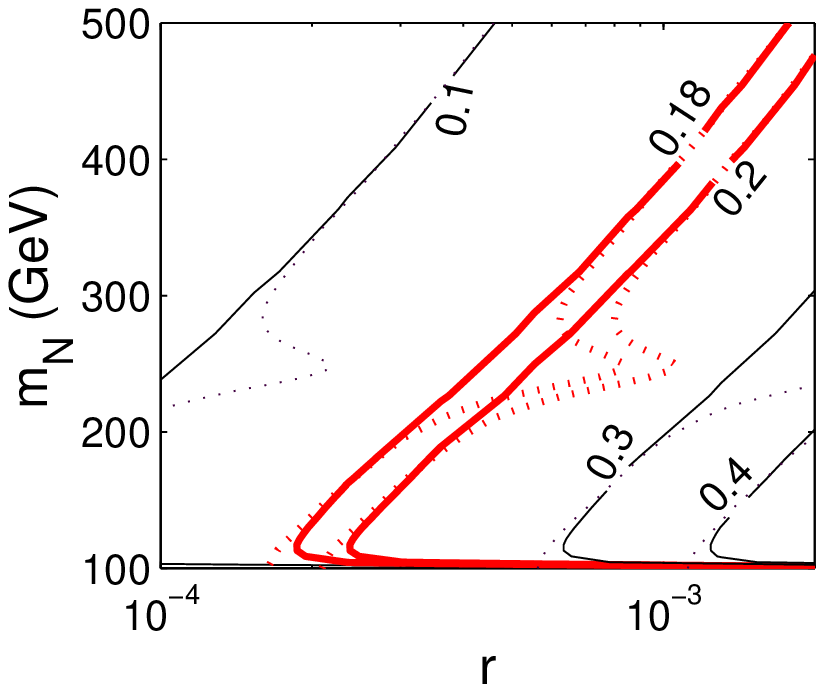}}
  \qquad\qquad
  \subfigure{\includegraphics[width=6.55cm]{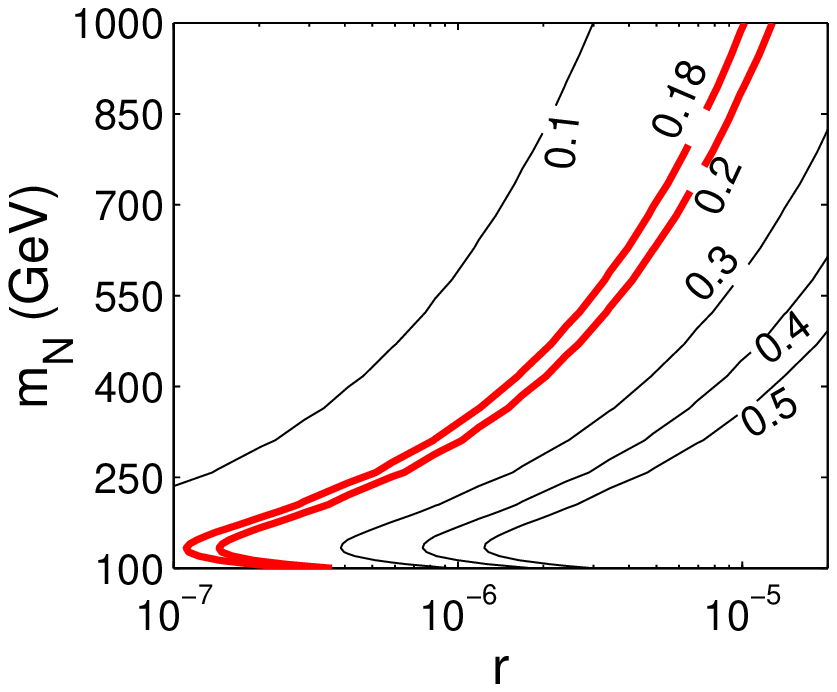}}
\caption{Shown are the constant $\Omega_N$-contours for the Dirac (left panel) and Majorana neutrinos (right panel) in the $(m_N,r)$-plane, where $m_N$ is the mass of the neutrino and $r$ is the ratio of the dark energy and radiation energy densities at $T=1$ MeV (see Eq.~(\ref{ecosmo2})). The area between the thick curves correspond to the observationally accepted region for the cold dark matter abundance~\cite{Spergel:2006hy}.}
\label{omegacontours}
\end{figure*}
where $K_i(y)$'s are modified Bessel functions of the second kind and $s$ is the usual Mandelstam invariant. For the total cross section $\sigma_{\rm tot}$ we included the $N\bar{N}$ annihilation to the final states $f\bar{f}$, $H^0H^0$ and $W^{+}W^{-}$, where $f$ is any fermion, $H^0$ the neutral Higgs boson and $W^{\pm}$ the charged vector bosons. We omitted annihilations to  technifermions because these rates would be just a small correcetion to already subleading fermionic channel. We did include the result of an effective, light ``SM-like" Higgs to illustrate qualitatively the behaviour near a mass threshold relevant for the underlying composite sector. In the Dirac case we used the complete expressions for these cross sections from ref.~\cite{Enqvist:1988we}. The dominant contribution to $\sigma_{\rm tot}$ at $m_N \gsim 100$ GeV comes from $Z$-mediated $s$-channel annihilation to longitudinal charged gauge bosons however, and the cross section for this sub-process is:
\begin{eqnarray}
  \label{ecosmo5}
  \sigma(N\bar{N}
  \rightarrow W^{+}_{L} W^{-}_{L})_D
   = \frac{G_{F}^{2}m_{W}^{4}\beta_{W}}{48\pi s \beta_N} \mid\hat D_Z\mid^{2}
      \times \phantom{hanna}
   \\ \nonumber
      \times (\hat{s}-2)^2 ( \hat{s}^2 - \hat{s}\hat m_N^2 
                                 - 4\hat{s} + 4\hat m_N^2 ),
\end{eqnarray}
where $\hat{s}= s/m_{W}^{2}$, $\hat m^2_N = m^2_N/m_{W}^{2}$, $\beta_N = (1-4m_N^{2}/s)^{1/2}$,
$\beta_W = (1-4m_{W}^{2}/s)^{1/2}$
and the $Z$ boson propagator factor is:
\begin{equation}
  \mid \hat D_Z\mid^2 = \frac{1}{(\hat{s}-\hat{m}_{Z}^{2})^{2}   
  + \hat \Gamma_Z^2 \hat m_Z^2},      
\end{equation}
with the scaled $Z$-boson width $\hat \Gamma_Z \equiv \Gamma_Z/m_{W}$. Together with the full cross section for the channel $\sigma(N\bar{N}\rightarrow f \bar{f})$ the approximation (\ref{ecosmo5}) is accurate to about a factor of two at $m_N \sim 100$ GeV. For larger masses the approximation becomes much better, reaching the order of a few per cents at $m_N\approx 1$ TeV. For the Majorana case the cross section is not available for the dominant $W^+W^-$-final state. We have calculated the Majorana-$N$ annihilation into longitudinal $W$-pairs, and the result is:
\begin{eqnarray}
  \sigma(N\bar N \rightarrow W^+_LW^-_L)_M
        =\frac{G_F^2m_W^4 \beta_W}{24\pi s}\times\\ \nonumber
\mid\hat{D_Z}\mid^{2}\hat{s}(\hat{s}-4) (\hat{s}-2)^2.
\end{eqnarray}
For the total cross section in the Majorana case we included also the annihilation to $f\bar f$-pairs, the cross section for which can again be found in ref.~\cite{Enqvist:1988we}.

Given the cross sections and the expansion history of the universe, we can solve Eqn.~(\ref{ecosmo1}) for $f(0)$, which gives the present ratio of $N$-number-density to the entropy density.  The fractional density parameter $\Omega_N$ can then be computed from 
\begin{eqnarray}
\label{ecosmo8}
\Omega_{N}\simeq 1.09 \times 10^6 \, a\,  m_N f(0)\,,
\end{eqnarray}
where $a=1$ for Dirac case and $a=1/2$ for Majorana neutrinos. As we explained in the introduction, our cosmological model mainly depends on two parameters, the neutrino mass $m_N$ and the dark energy to radiation energy density ratio at one MeV temperature $r$ . In Fig.~\ref{omegacontours} we show the constant $\Omega_{N}$-contours in the $(r,m_N)$-plane resulting from our computations. The thick lines mark the boundaries of the most probable region of the dark matter abundance $\Omega_N \approx 0.18-0.2$~\cite{Spergel:2006hy}. Solid lines correspond to the  Higgs boson mass $m_{H^0}=100$ GeV and the dashed ones to $m_{H^0} = 500$ GeV. The dip in dashed contours in the Dirac case is thus result of the $s$-channel pole in the $W^+W^-$ production via a Higgs-exchange (this interaction channel was not included in the Majorana case). Note that the effect of the Higgs interaction is neglible away from the pole $m_N \simeq m_{H^0}/2$.

Since $\Omega_N \sim \langle v\sigma\rangle/H \sim \langle v\sigma\rangle/\sqrt{r}$, an underestimate by a factor $b$ in the cross section would lead to a bias towards smaller $r$ of a factor $b^2$ in the position of the $\Omega_N$-contours in Fig.\ref{omegacontours}.  Assuming that the longitudinal final state approximation is of similar accuracy for both Dirac and Majorana cases, we estimate that our calculation for $\sigma_{{\rm tot},M}$ may be off by a factor of two at the worst case when $m_N \sim 100$ GeV. Thus the Majorana contours may be wrong (too much to the left, {\it i.e.}~too ``optimistic") in $r$ by a factor of four at $m_N \sim 100$ GeV, but the error should soon become neglible at larger $m_N$.  This uncertainty does not change our results qualitatively however: both in the Dirac and the Majorana cases a large enough relic density of neutrinos can be produced with a dark energy component that is neglible during nuclosynthesis.

Let us now comment on how important the modified expansion history due to kination is to our results. This is best seen by comparing the relative size of the conventional term $h$ to the kination factor $(x_{\rm f}/x_0)^2\,r$ in the Hubble parameter (\ref{ecosmo2}) during the freeze-out. For example for a Dirac neutrino with $m_N=150$ GeV the freeze-out occurs at $T_{\rm f} \approx 6.6$ GeV, and the kination factor turns out to be very large: $(x_{\rm f}/x_0)^2\,r/h \approx 10^5$. A modified expansion history is essential here, but it should be stressed that no observational evidence exists to constrain even this large deviation from a standard expansion history prior to BBN. In the Majorana case with the same mass the necessary speed-up factor is rather modest however. The freeze-out now occurs at $T_{\rm f} \approx 6.3$ GeV and one finds a kination factor $(x_{\rm f}/x_0)^2r/h \approx 50$. Because $\Omega_N \sim \langle v\sigma\rangle/\sqrt{r}$, the same relic density could have been obtained in a standard expansion history if the cross section was smaller by a factor $\sim 10$. It might be possible to arrange this in an extension of our model, where $N$ is taken to be a mainly right-handed Majorana state.

We also calculated the limits for the density parameter from the direct dark matter searches. The relevant cross sections for the nuclear recoil experiments can be found for example from~\cite{Rybka:2005vv}. For Dirac neutrinos scattering coherently from a nucleus via neutral current the spin-independent cross section is
\begin{eqnarray}
\label{ecosmo9}
  \frac{d\sigma}{dQ} = \frac{G_F^2 m_A}{16 \pi v^2}
   [Z(1-4\sin^2\theta_W) - N ]^2\times  \\ \nonumber
     \times \exp(-2m_AQR^{2}/3),
\end{eqnarray}
where $Q$ is recoil energy, $m_A$ is the mass of the nucleus of the target atom with $A = N+Z$ nucleons ($Z$ protons and $N$ neutrons), $v$ is the velocity of the neutrino relative to detector and $R \approx 1.2 \, A^{1/3}$ fm. The cross section for Majorana neutrinos is
\begin{eqnarray}
  \frac{d\sigma}{dQ}= \frac{G_F^2 m_A}{\pi v^2}C^2\lambda^2J(J+1),
  \label{ecosmo10}
\end{eqnarray}
where the factor $C^2$ is related to the quark spin content of the nucleon and $\lambda^{2}J(J+1)$ is related to the nucleon spin content of the nucleus. We used the value $C^{2}\lambda^{2}J(J+1)=0.026$ from~\cite{Lewin:1996}. We included also the interaction through an effective Higgs exchange taking SM-result as a guide. This cross section is same for Dirac and Majorana neutrinos \cite{Rybka:2005vv}:
\begin{eqnarray}
\label{ecosmo11}
  \frac{d\sigma}{dQ}= \frac{G_F^2 m_A}{8 \pi v^2}
           \frac{m_N^2m_A^2}{m_{H^0}^4}\exp(-2m_A QR^2/3),
\end{eqnarray}
where $m_{H^0}$ is the mass of the Higgs boson.
For the Dirac case (\ref{ecosmo11}) is neglible compared to interaction (\ref{ecosmo9}). However, in the Majorana case it dominates the detection sensitivity at $m_N \gsim 250$ GeV.
The total count rate seen in the detector over a recoil energy range $\Delta Q$ is~\cite{Jungman:1995df}
\begin{equation}
  R = \Delta Q \frac{\rho}{m_N m_A}   
      \int_{v_{\rm min}}^{v_{\rm max}}\frac{d\sigma}{dQ}v f_1(v){\rm d}v \,,
\label{ecosmo12}
\end{equation}
where $f_1(v)$ is the velocity distribution of neutrinos in the earth frame  
\begin{eqnarray}
\label{ecosmo13}
  f_1(v){\rm d}v &=& \frac{v}{\sqrt{\pi}v_0 v_e}
         \Big[\exp \Big(-\frac{(v-v_e)^2}{v_0^2} \Big)
\\ \nonumber && \phantom{hanni.}
         - \exp \Big(-\frac{(v+v_e)^2}{v_0^2} \Big)\Big]\, {\rm d}v \,,
\end{eqnarray}
corresponding to a Maxwellian distribution in galaxy frame. Velocities of the Sun and the Earth around the galaxy center are taken to be $v_0=220$ km/s and $v_e=1.05v_0$ respectively~\cite{Jungman:1995df}. 
\begin{figure*}[t]
\centering
  \subfigure{\includegraphics[width=6.6cm]{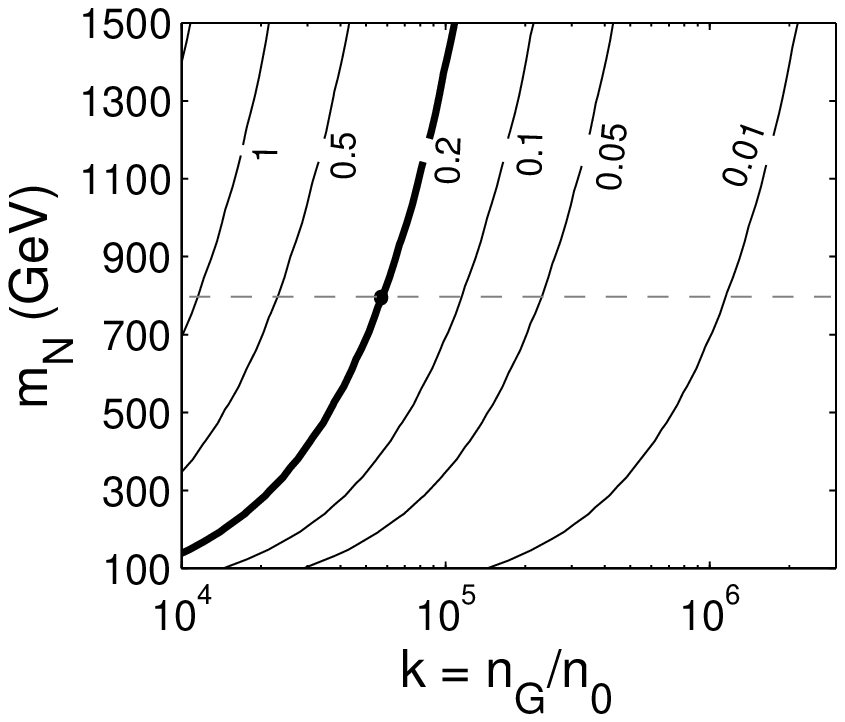}}
  \qquad\qquad
  \subfigure{\includegraphics[width=6.47cm]{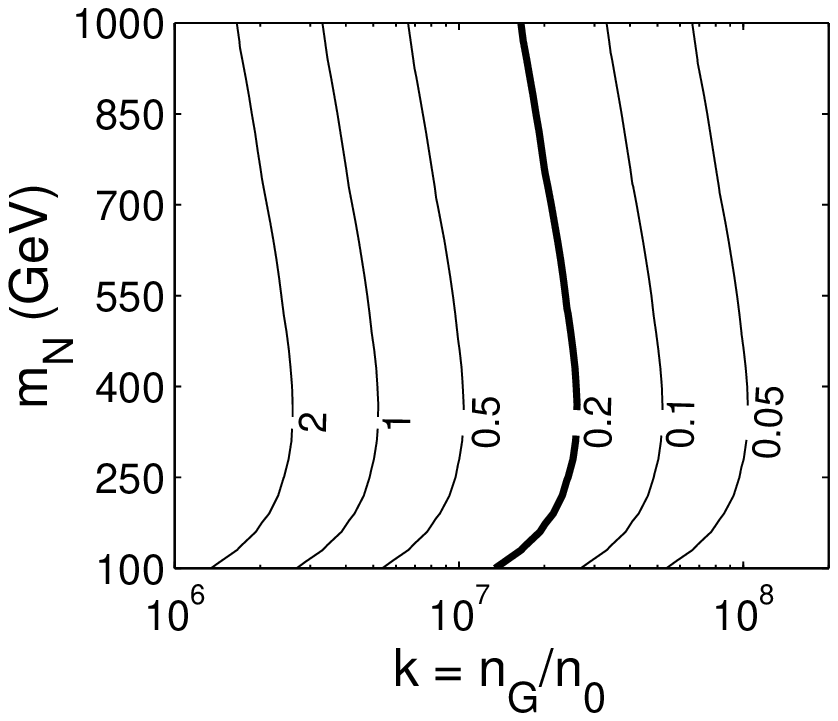}}
\caption{Shown are the contours of constant $\Omega_N$ which lead to a limiting detector sensitivity for a given mass $m_N$ and a local enhancement factor $k$.  We used the Higgs mass $m_{H^0}=200$ GeV for the rate (\ref{ecosmo11}). For a given contour the area to the left from the curve is allowed. {\em Left panel}: Dirac case. Dashed line indicates the $k=1.2\times 10^4/\Omega_N$-envelope, the region above which is allowed given local dark matter density $\rho_G=0.06 \, \rm GeV/cm^3$. {\em Right panel} Majorana case.}
\label{lablimits}
\end{figure*}
Following ref.~\cite{Rybka:2005vv} we calculated our limits using the first data bin above the detector threshold; this is the most sensitive one because the recoil spectrum falls rapidly as a function of increasing recoil energy. The upper limit on number of events seen in this data bin then gives our upper limit for the local neutrino density. Using the fact that no counts were observed above the recoil energy $Q=20$keV with the energy resolution $\Delta Q=1.5$ keV and exposure time $\tau^{\star}=19.4$ kg-days in the CDMSII experiment~\cite{Akerib:2004fq}, we obtain our 90$\%$ confidence limits for neutrino density as follows~\cite{Rybka:2005vv}:
\begin{eqnarray}
  \rho < N_{90}\Big(\frac{\tau^{\star}\Delta Q}{m_N m_A} 
       \int_{v_{\rm min}}^{v_{\rm max}} 
               \frac{d\sigma}{dQ} v f_1(v) {\rm d}v\Big)^{-1},
\label{ecosmo14}
\end{eqnarray}
where the 90$\%$ confidence count limit $N_{90} = 2.3$ was used corresponding to an expected background of 0.3 events in the first bin~\cite{Akerib:2004fq}. Note that the limit coming from (\ref{ecosmo14}) does not directly constrain the cosmological density $\Omega_N$. Rather, a limit on cosmological density can be obtained only after the ratio $k \equiv n_G/n_0$ between local ($n_G$) and cosmological ($n_0$) WIMP densities is determined. 
However, even the local dark matter density is not very well known, and although most evaluations give a value $\rho_{\rm G} \approx 0.3-0.4 \; \rm GeV/cm^3$~\cite{Jungman:1995df}, a density as low as $\rho_G \approx 0.06 \; \rm GeV/cm^3$ might be acceptable~\cite{Jungman:1995df}. For a model with a cosmological density $\Omega_N$, we then must find solutions with {\em an acceptable} clustering factor that falls into the range $k \approx (1.2-12)\times 10^4/\Omega_N$. 
We present the experimental constraints coming from (\ref{ecosmo14}) in Fig.~\ref{lablimits}. Contours show the limiting clustering factor $k(m_N,\Omega_N)$ for which the observational limit is saturated; the region to the right of each contour is excluded.
In the Dirac case (the left panel) a solution with $\Omega_N \approx 0.2$ is allowed for $m_N \gsim 800$ GeV if one uses the least restrictive clustering value $k_{\rm min} \approx 6 \times 10^4$. This is quite unlikely case and a pure Dirac neutrino is probably a poor dark matter candidate in our model. In the Majorana case (right panel) the bound is rather weak however, and in particular for a neutrino with a mass on the target range $m_N \sim 100-500$ GeV (from the precision constraints, see Fig.(\ref{majorana_leptons})) the experimental bound is easily evaded with $\Omega_N\approx 0.2$ even with a conservative clustering value $k\approx 6 \times 10^5$. We then conclude that a pure left chiral Majorana neutrino is a good dark matter candidate in our model. 

\section{Conclusions} 
 
We have considered a possibility that the dark matter observed in the universe could be a massive fourth family neutrino predicted by a recently discovered technicolor model for the electroweak symmetry breaking~\cite{Sannino:2003xe, Sannino:2004qp}. This neutrino can be taken to be either a Dirac or Majorana  particle, and in both cases it can be the lighter member of the new lepton doublet. Its is predicted to be of order of a few hundred GeV by the constraints coming from electroweak precision experiments. The relic density of such a neutrino can fall to the observationally determined range $\Omega_N \approx 0.2$ if the universe has a nonstandard expansion history including an early dominance by a rolling scalar field, as predicted by many models for dynamical dark energy.  Moreover, if the new neutrino is a Majorana state, it could easily have escaped detection in the direct dark matter searches. Since the necessary speed-up factor for the expansion of the universe during the freeze-out is only about a factor 6-10 in the Majorana case, it might be possible to make the model consistent with the observation under standard expansion history if the dark matter neutrino was mainly an inert right-handed Majorana state. However, as one needs at least one left handed doublet in the model to cure the Witten anomaly, this would imply an extension of the particle content of the theory.
Another issue worth a further study would be to work out the natural mixing patterns in the neutrino sector including the heavy state and see how the necessary (near) stability of $N$ might arise from these considerations.

%%%%%%%%%%%%%%%%%%%%%%%%%%%%%%%%%%%%%%%%%%%%%%%%%%%%%%%%%%%%%%%%%%%%%%%%%%%%%

\section*{Acknowledgments}

\noindent We thank F.~Sannino for discussions and for careful reading of this manuscript.

%%%%%%%%%%%%%%%%%%%%%%%%%%%%%%%%%%%%%%%%%%%%%%%%%%%%%%%%%%%%%%%%%%%%%%%%%%%%%

%%%%%%%%%%%%%%%%%%%%%%%%%%%%%%%%%%%%%%%%%%%%%%%%%%%%%%%%%%%%%%%%%%%%%%%%%%%%%

\end{document}